\documentclass[graybox]{svmult}

\usepackage{mathptmx}       
\usepackage{helvet}         
\usepackage{courier}        
\usepackage{type1cm}        
\usepackage{makeidx}         
\usepackage{graphicx}        
\usepackage{multicol}        
\usepackage[bottom]{footmisc}

\makeindex             

\begin{document}

\title*{Dynamics of apparent horizons in quantum gravitational collapse}
\author{Yaser Tavakoli, Andrea Dapor and Jo\~ao Marto}
\institute{Yaser Tavakoli \at Departamento de F\'{\i}sica, Universidade da Beira Interior,
 6200 Covilh\~a, Portugal \email{tavakoli@ubi.pt}
\and Andrea Dapor \at Instytut Fizyki Teoretycznej, Uniwersytet Warszawski,
ul. Ho\.{z}a 69, 00-681 Warsaw, Poland \email{andrea.dapor@fuw.edu.pl}
\and Jo\~ao Marto \at Departamento de F\'{\i}sica, Universidade da Beira Interior,
 6200 Covilh\~a, Portugal \email{jmarto@ubi.pt}}
\maketitle

\abstract{ We study the gravitational collapse of a massless scalar field within the effective scenario of loop quantum gravity.
Classical singularity is avoided and replaced by a quantum bounce in this model.
It is shown that, quantum gravity effects predict a threshold scale below which no horizon can form as the collapse evolves towards the bounce. 
}

\section{Gravitational collapse with a massless scalar field}
\label{sec:1}

Loop quantum gravity (LQG) provides a fruitful ground to investigate the resolution of the classical singularities which arise in the gravitational collapse. In view of this approach, it is of interest and well motivated to further assess how LQG can affect on the evolution
of the trapped surfaces and the horizon formation during the collapse.
In this paper we introduce a spherically symmetric framework for the gravitational collapse.
We consider the dynamical space-time inside the collapsing sphere to be \emph{homogeneous} and is given by the flat Friedmann-Robertson-Walker (FRW) metric as \cite{us},
\begin{equation}
g^-_{ab}dx^adx^b=-dt^2+a^2(t)dr^2+R^2(t,r)d\Omega^2,
\label{metric}
\end{equation}
where $R(t,r)=ra(t)$ is the area radius of the collapse.
In terms of the $SU(2)$ variable of LQG we introduce the phase space variables for the interior space-time to be $c = \gamma \dot{a}$ and  $p=a^2$ \cite{A-B-L}.
Considering a \emph{homogeneous} and \emph{massless} scalar field for the collapsing matter source, the four dimensional phase space $(c,p; \phi, p_{\phi})$ is governed by the fundamental Poisson brackets $\{c,p\}=(8\pi G/3)\gamma$, and $\{\phi,p_{\phi}\}=1$.
Then, the total classical Hamiltonian constraint for the system can be written as \cite{A-B-L}
\begin{equation}
C=C_{\mathrm{gr}}+C_{\phi}=-(6/\gamma^2)c^2\sqrt{|p|}+8\pi Gp_{\phi}^2/|p|^{3/2}=0,
\label{Hconstraint}
\end{equation}
where $p_{\phi}$ is a constant of motion in the classical theory.
In addition, the energy density and pressure of the collapsing matter reads
\begin{equation}
\rho_\phi=p_\phi^2/2V^2=p_\phi^2/2|p|^{3}= -p_\phi.
\label{energy2}
\end{equation}
The point $p=0$ characterizes a situation where the volume $V=|p|^{3/2}$, of the collapsing matter is zero and the energy density of the matter cloud diverges. Therefore, if this point lies on any dynamical trajectory, it is an end point of that trajectory corresponding to a curvature singularity which characterizes the collapse end state.

Interior space-time should be matched at the boundary to a sufficient exterior geometry such as 
the generalized Vaidya metric (c.f. see Ref. \cite{us}):
\begin{equation}
g_{ab}^{+}dx^adx^b=-(1-2M/r_\mathrm{v})d\mathrm{v}^2-2d\mathrm{v}dr_{\mathrm{v}}+r_\mathrm{v}^2d\Omega^2. 
\label{metric2}
\end{equation}
The formation or avoidance of a black hole in the exterior region depends on developing an apparent horizon in the interior space-time.
Therefore, in this paper we are mainly concerned with whether or not horizons can form during the evolution of the
interior space-time in the presence of the quantum effects (c.f. section \ref{sec:2}).

In order to study the geometry of trapped surfaces inside the star, let us introduce the radial null geodesics satisfying $g_{ab}^{-}dx^adx^b=0$, 
whose expansions, $\theta_{\pm}$, measure whether the bundle of null rays normal to the collapsing sphere is diverging $(\theta_{\pm}>0)$ or converging
$(\theta_{\pm}<0)$. Let us define the useful parameter $\Theta\equiv \theta_{+}\theta_{-}$ \cite{Hayward}:
\begin{equation}
\Theta=\left(\dot{R}^2/R^2-1/R^2\right)/2 .
\label{theta1-q}
\end{equation}
So that, the space-time is referred to, respectively, as trapped, untrapped and marginally trapped if:
$ \Theta(t,r)>0,~ \Theta(t,r)<0$, and  $\Theta(t,r)=0$.

\section{Quantum geometry of trapped surfaces}
\label{sec:2}

%
\begin{figure}
\includegraphics[height=1in]{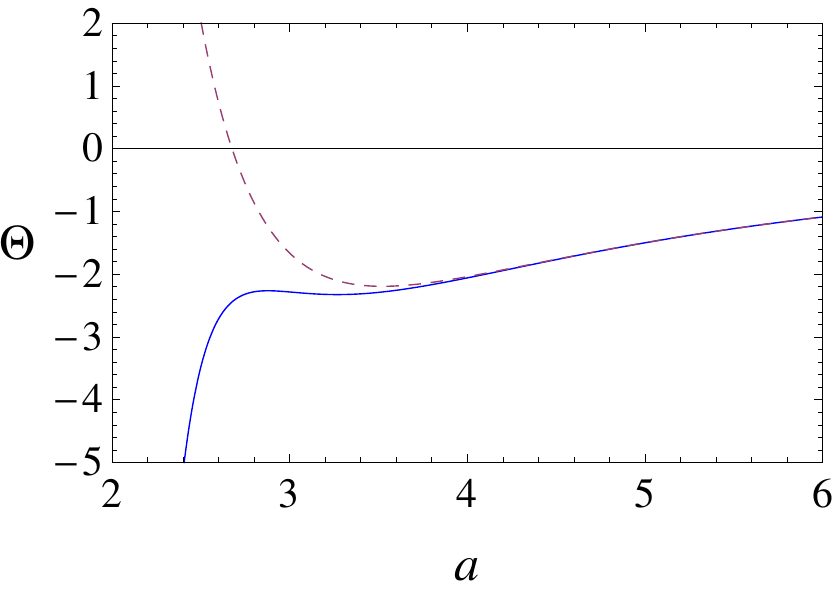}\quad{}\includegraphics[height=1in]{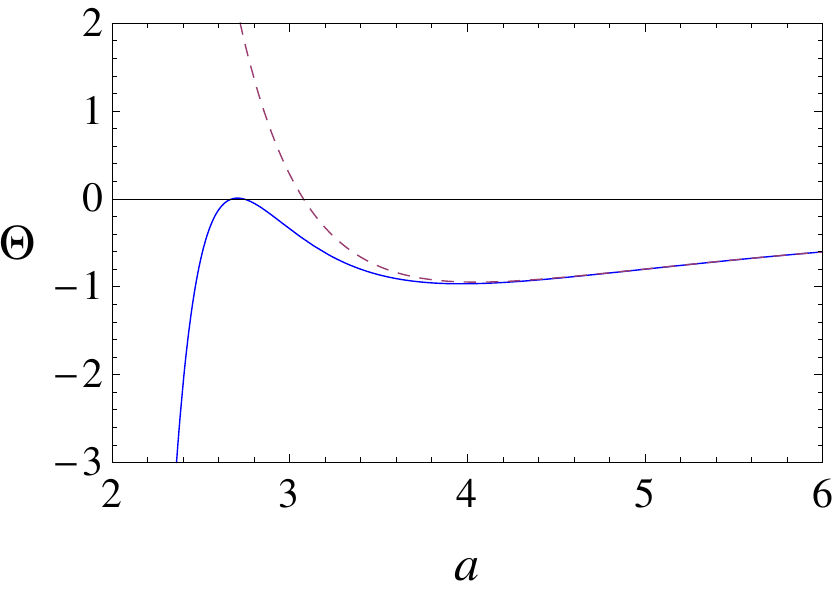}\quad{}\includegraphics[height=1in]{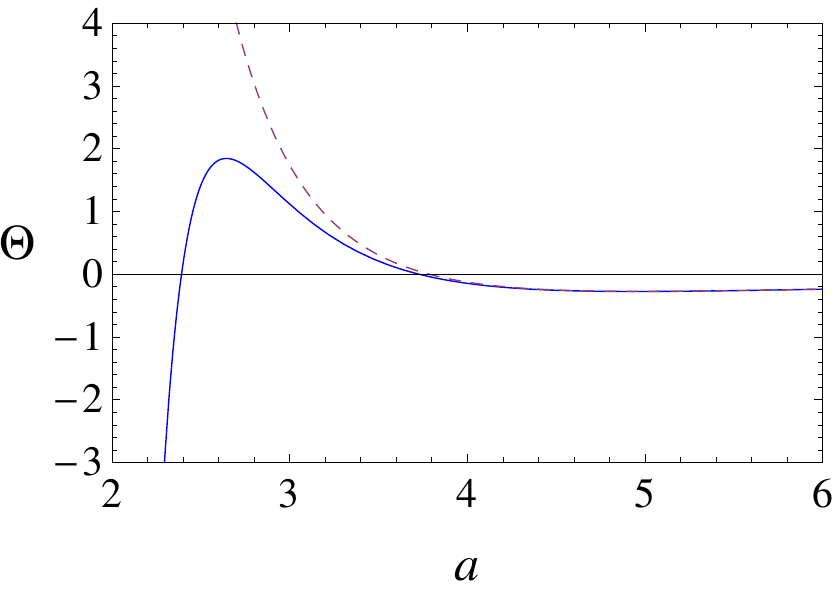}
\caption{{\footnotesize Behaviors of $\Theta(a)$ in the classical (dashed curve) and quantum (solid curve) regimes for the value of parameter $G=c_{\mathrm{light}}=1$, $p_\phi=10~000$, and the different values of $r_\mathrm{b}$.}}
\label{F2}
\end{figure}
%

The strategy here is to consider the interior space-time as a classical phase space, of the FRW model, coupled to matter that is equipped with a quantum corrected Hamiltonian constraint; the effective scenario of LQG. This constraint and its resulting Hamilton equations of motion, provide the full set of effective Einstein's equation for the collapsing model herein. In this quantization scheme, the curvature 2-form $c^2$ is modified by $SU(2)$ holonomy along suitable loops; the leading order quantum corrections  are captured in the following effective Hamiltonian \cite{A-P-S}:
\begin{equation}
C_{\mathrm{eff}}=-(3/8\pi G\gamma^2\mu_o^2)\sqrt{|p|}\sin^2(\mu_oc)+ B(p)p_{\phi}^2/2,
\end{equation}
where $B(p)$ is the eigenvalues of operator $\widehat{|p|^{-3/2}}$ \cite{A-P-S}.
Consider a new variable $v$, given as the eigenvalue of the volume operator
$\hat{V}|v\rangle=(8\pi \gamma/6)^{\frac{3}{2}}K^{-1}|v|\ell_{\mathrm{Pl}}^3|v\rangle$: then,
equations of motion can be obtained by using the Hamilton's equation \cite{A-P-S}:
\begin{eqnarray}
& \dot{v} = \{v, C_{\mathrm{eff}}\}=  (2|v|^{1/3}/\gamma\mu_o K)(8\pi \gamma\ell_{\mathrm{Pl}}^2/6)^{1/2}\sin(\mu_o c)\cos(\mu_o c),  \label{Hamilton1}\\
& \dot{\phi} = (1/16\pi G)\{\phi, C_{\mathrm{eff}}\} = (8\pi \gamma\ell_{\mathrm{Pl}}^2/6)^{-3/2}Kp_\phi|v|^{-1}\ .
\label{Hamilton2}
\end{eqnarray}
where $K=(2\sqrt{2}/3\sqrt{3\sqrt{3}})$. Using $C_{\mathrm{eff}}=0$ in Eqs. (\ref{Hamilton1}) and (\ref{Hamilton2}), the modified Friedmann equation reads
\begin{equation}
\dot{a}^2/a^2=\dot{v}^2/(9v^2)=(8\pi G/3)\rho_\phi(1-\rho_\phi/\rho_{\mathrm{cr}}),
\label{Friedman-eff}
\end{equation}
where $\rho_{\mathrm{cr}}=\sqrt{3}/(16\pi^2\gamma^3G^2\hbar)\approx 0.82\rho_{\mathrm{Pl}}$.
Notice that, in the effective scenario of LQG, the energy density $\rho_\phi= p_\phi^2/2|p|^3$
holds in the range $\rho_i<\rho_\phi<\rho_{\mathrm{cr}}$. 
In the limit $\rho_\phi\ll \rho_{\mathrm{cr}}$, the standard classical Friedmann equation, $\dot{a}^2/a^2=(8\pi G/3)\rho_\phi$, is recovered.
Equation (\ref{Friedman-eff}) shows that, the speed of collapse $|\dot{a}|$ starts to increase initially and reaches its maximum when $\rho_\phi=0.4\rho_\mathrm{cr}$:
\begin{equation}
|\dot{a}|_{\mathrm{max}}\ =\ \sqrt{4\pi G/5}(0.4\rho_\mathrm{cr})^{1/3}p_\phi \ ,
\end{equation}
at the scale $a_{\mathrm{max}}$, given by $a_{\mathrm{max}}\ \approx\ 1.16~ a_{\mathrm{cr}}$, where  
$a_{\mathrm{cr}}:=(p_\phi^2/2\rho_\mathrm{cr})^{1/6}$.
Hereafter, $|\dot{a}|$ decreases by time and vanishes as $\rho_{\phi}\rightarrow\rho_{\mathrm{cr}}$; thus,
the collapsing star bounces at the minimum volume $V_{\mathrm{min}}=a_{\mathrm{cr}}^3\approx (1.3\times10^{-33}\mathrm{cm}^3)p_\phi$.
\begin{figure}
\includegraphics[height=1.4in]{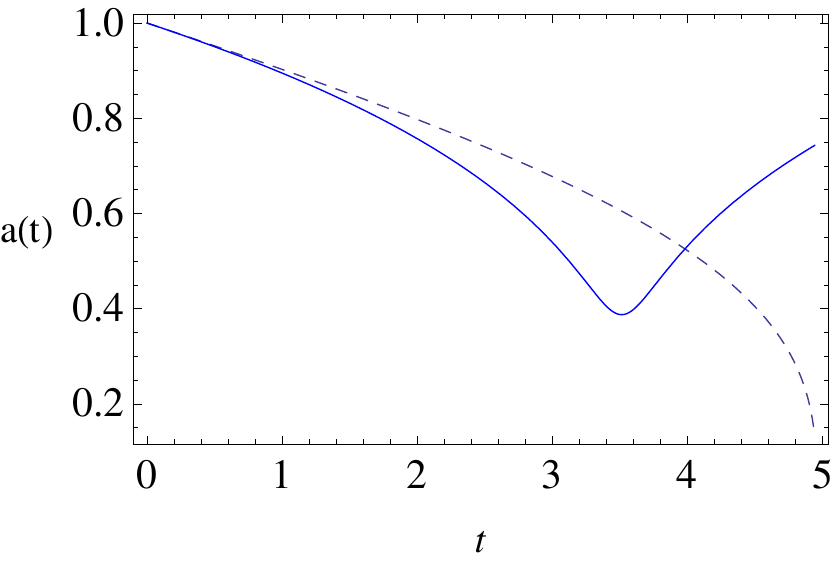}\quad{}\includegraphics[height=1.4in]{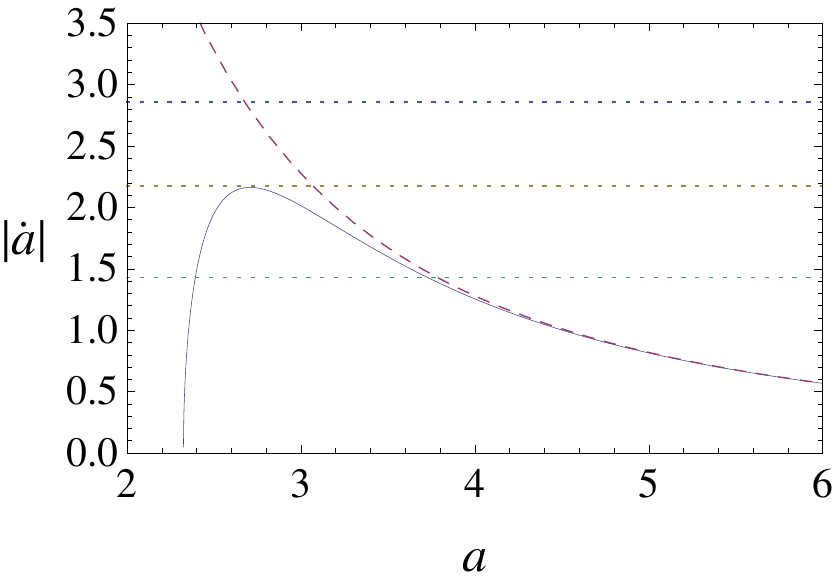}
\caption{{\footnotesize Left figure shows behavior of the scale factor, $a(t)$, with time in classical (dashed curve) and semiclassical (solid curve) regimes. The right plot shows the speed of collapse, $|\dot{a}|$, with respect to the scale factor $a$, in classical (dashed curve) and semiclassical (solid curve) regimes. The horizontal dotted lines correspond to different values of $r_\mathrm{b}$; the upper, middle and the lower lines are, respectively, the cases $r_\mathrm{b}<r_\mathrm{th}$,~$r_\mathrm{b}=r_\mathrm{th}$, and $r_\mathrm{b}>r_\mathrm{th}$. }}
\label{F1}
\end{figure}

To discuss the quantum geometry of the  trapped region from the perspective of the effective scenario of LQG,
it is convenient to rewrite  $\Theta$ in Eq. (\ref{theta1-q}) as \cite{us}
\begin{equation}
\Theta (a)
= 4\pi G/3\rho_\phi(1-\rho_\phi/\rho_{\mathrm{cr}})-1/(2r_{\mathrm{b}}^2a^2) ,
\label{theta2-q2}
\end{equation}
where we have replaced $\dot{a}^2/a^2$ here by the effective Hubble rate Eq. (\ref{Friedman-eff}), and $\rho_\phi=p_\phi^2/(2a^6)$.
Figure \ref{F2} shows the behavior of $\Theta$ against the scale factor $a$ for the different choices of the initial conditions for constants
$r_\mathrm{b}$ and $p_\phi$. Equation of apparent horizon on the effective geometry can be obtained by setting $\Theta=0$.
The left plot in Fig. \ref{F2} (solid curve) indicates an untrapped interior space-time without any horizon forming,
whereas two others show trapped regions; the middle and the right plots correspond to one and two horizons forming, respectively.
Notice that,  only one horizon would always form classically (dashed curves in Fig. \ref{F2}).

Using Eq. (\ref{theta1-q}), we can determine the speed of the collapse at which horizon can form, where $\Theta=0$, from which we get
$\dot{R}^2=1$, and whence, $|\dot{a}|_{\mathrm{AH}} =1/r_\mathrm{b}$.
When the speed of collapse, $|\dot{a}|$, reaches the value $1/r_\mathrm{b}$,
apparent horizons form (Fig. \ref{F1}).
In other words, in order to have the horizon formation, $r_\mathrm{b}$ must satisfy $r_\mathrm{b}\geq |\dot{a}|_{\mathrm{max}}^{-1} $.
It is convenient to introduce a radius $r_\mathrm{th}$ as,
\begin{equation}
r_\mathrm{th}\ :=\  |\dot{a}|_{\mathrm{max}}^{-1} .  
\end{equation}
Therefore, as it is shown in Fig. \ref{F1}, radius $r_\mathrm{th}$ is a threshold radius for the horizon formation. More precisely, for the case $r_{\mathrm{b}}\ < \ r_\mathrm{th}$, no horizon would form as collapse evolves; the case $r_{\mathrm{b}}\ =\ r_\mathrm{th}$,
corresponds to the formation of a dynamical horizon at the boundary of two regions; and finally, for the case
$r_{\mathrm{b}}\ >\ r_\mathrm{th}$, two horizons will form, one inside and the other outside of the collapsing matter.
%


%

\begin{acknowledgement}
The authors would like to thank R. Goswami and J. Velhinho for the useful discussion and suggestions.
YT is supported by the FCT (Portugal) 
through the fellowship SFRH/BD/43709/2008. 
\end{acknowledgement}

\end{document}